# Polarization-Independent Wavelength Demultiplexer Based on Single Etched Diffraction Grating Device

**CHENGUANG LI, BO XIONG, AND TAO CHU***

*College of Information Science and Electronic Engineering, Zhejiang University, Hangzhou, 310027, China*
**Corresponding author: chutao@zju.edu.cn*

**Abstract:** Polarization-compatible receivers are indispensable in transceivers used for wavelength division multiplexing (WDM) optical communications, as light polarization is unpredictable after transmission through network fibers. However, the strong waveguide birefringence makes it difficult to realize a polarization-independent wavelength demultiplexer in a silicon photonic (SiPh) receiver. Here, we utilized the birefringence effect for simultaneously demultiplexing wavelengths and polarizations, and experimentally demonstrated a novel polarization-independent wavelength demultiplexer with a single device on a SiPh platform. The principle was validated on an etched diffraction grating (EDG), which successfully split the arbitrarily polarized light containing four wavelengths into eight channels with single-polarization and single-wavelength signals. Polarization-dependent losses of 0.5–1.8 dB, minimum insertion loss of 0.5 dB, and crosstalks lower than −30 dB were experimentally measured. Thus, a promising general solution was developed for implementing polarization-independent WDM receivers and other polarization-independent devices on SiPh and other platforms with birefringent waveguide devices.

## 1. INTRODUCTION

Wavelength division multiplexing (WDM) is one of the most important technologies for optical communications with its outstanding advantages in terms of low cost and high capacity [1,2], which overcomes the transmission limit of a single optical fiber by carrying signals with multi-wavelength light [3,4]. Wavelength multiplexers and demultiplexers combine and separate multi-wavelength lights, respectively, and are the key devices of WDM systems. However, owing to the change in the state of polarization (SOP) of light in the network fibers, the demultiplexers must be able to operate in arbitrary SOPs to avoid the serious problem of polarization-dependent loss (PDL) [5,6]. Usually, polarization-independent demultiplexers can be easily fabricated on a $SiO_2$-based planar light circuit (PLC) platform; unfortunately, such devices have large footprints and cannot be integrated with photodiodes (PDs) monolithically, leading to a significant reduction in integration density [7-9]. In contrast, InP-based demultiplexers can be monolithically integrated with PDs and are polarization independent [10,11]. However, they still suffer from large footprints and high costs. Lately, silicon photonic (SiPh) integration is being considered a very promising platform for constructing WDM systems with the benefits of high-density monolithic integration, low cost, and compatibility with complementary metal-oxide-semiconductor (CMOS) processing [12]. Regretfully, owing to the significant structural birefringence in silicon-based waveguides, the operation of demultiplexers on a SiPh platform is usually restricted to a single polarization [13-19].

Complicated polarization diversity schemes are necessary to avoid the polarization dependence of devices on a silicon-based platform, such as SiPh demultiplexers [20-24]. In this system, light with arbitrary SOPs is divided into TE (transverse electric) and TM (transverse magnetic) polarizations; then, they are converted to the same TE/TM polarizations and transmitted separately to two similar sets of light circuits for processing. Finally, they are converted back to different TE/TM polarizations for combination and output, which is obviously very complicated, as it needs two similar sets of light circuits for signal processing. Recently, researchers demonstrated polarization-independent demultiplexers by cascading a single demultiplexer with polarization-handling devices such as a polarization beam splitter (PBS) [25,26] or a series of polarization rotators (PRs) [27,28]. In these devices, the polarization dispersions are compensated with the angular dispersion via PBSs or polarization rotation via PRs. However, the cascaded devices inevitably have larger losses and footprints. Other designs of polarization-independent demultiplexers are as follows. One is based on a thick-top-silicon based platform [29], in which high-speed modulators and high-density integration are difficult to realize owing to the large waveguide section size. Further, a nanostructured free propagation region (FPR) has been used to compensate for the polarization dispersion, which also causes additional loss and difficulties in fabrication [30,31]. Thus, handling the received light beams with arbitrary SOPs for wavelength demultiplexing remains an unsolved serious issue for SiPh devices; it requires a separate large silica PLC demultiplexer chip, which has to be packaged in a commercial transceiver along with monolithic integrated SiPh chips [8,9].

Here, we propose a novel polarization-independent wavelength demultiplexer based on a single SiPh etched diffraction grating (EDG) device. By utilizing the difference in the effective refractive index between the TE and TM polarizations in high-index-contrasted slab waveguides, lights with various polarizations and wavelengths can be transmitted to different output channels on the EDG Roland circle. Thus, the demultiplexing of the wavelengths and polarizations can be realized in a single EDG device simultaneously. Based on this idea, a polarization-independent EDG demultiplexer for a coarse WDM (CWDM) was fabricated and verified on a silicon nitride ($Si_3N_4$) thin-film platform. The measurement results showed that the EDG had insertion losses ranging from 0.5 dB to 2.4 dB, crosstalks below −30 dB, and PDLs between 0.5–1.8 dB at four target wavelengths. Moreover, a scan of the SOP of the incident light revealed that the outputs at four wavelengths remained almost stable with small standard deviations of 0.4–0.5 dB for the equator and 0.2–0.3 dB for the longitude of the Poincaré sphere. These results prove that a polarization-independent wavelength demultiplexer with low loss was successfully realized with a single EDG device. This work provides a general solution for polarization-dependent WDM transceivers in SiPh and other platforms with birefringent waveguide devices.

## 2. PRINCIPLE AND DESIGN

The schematic of the proposed polarization-independent EDG demultiplexer is shown in Fig. 1(a). An EDG performs wavelength multiplexing and demultiplexing by using the phase difference induced by the wavelength-dependent effective refractive index. Owing to the internal birefringence, the effective refractive indices of the TE and TM polarizations usually have significant differences, which result in EDG polarization dependence. However, in our design, this polarization dispersion was utilized as a new degree of freedom to build a multi-dimensional demultiplexer useful for both polarizations and wavelengths. By elaborately designing the polarization dispersion to achieve the phase matching of the EDG in the polarization dimension, light beams with different polarizations can be separated in the Roland circle to realize polarization demultiplexing based on wavelength demultiplexing. The proposed EDG device is fabricated on a $Si_3N_4$ thin-film platform consisting of four main components: input waveguides, FPR, etched gratings, and TE/TM output waveguides. When light with multi-wavelengths in arbitrary SOPs is input into the specially designed EDG, it will be demultiplexed into multiple light beams with single polarization (TE/TM) and single wavelength, and the outputs will be generated in different channels depending on the polarization and wavelength. By feeding the output TE and TM light beams of the same wavelength into one PD, wavelength demultiplexing is achieved with polarization independence in a single EDG device.

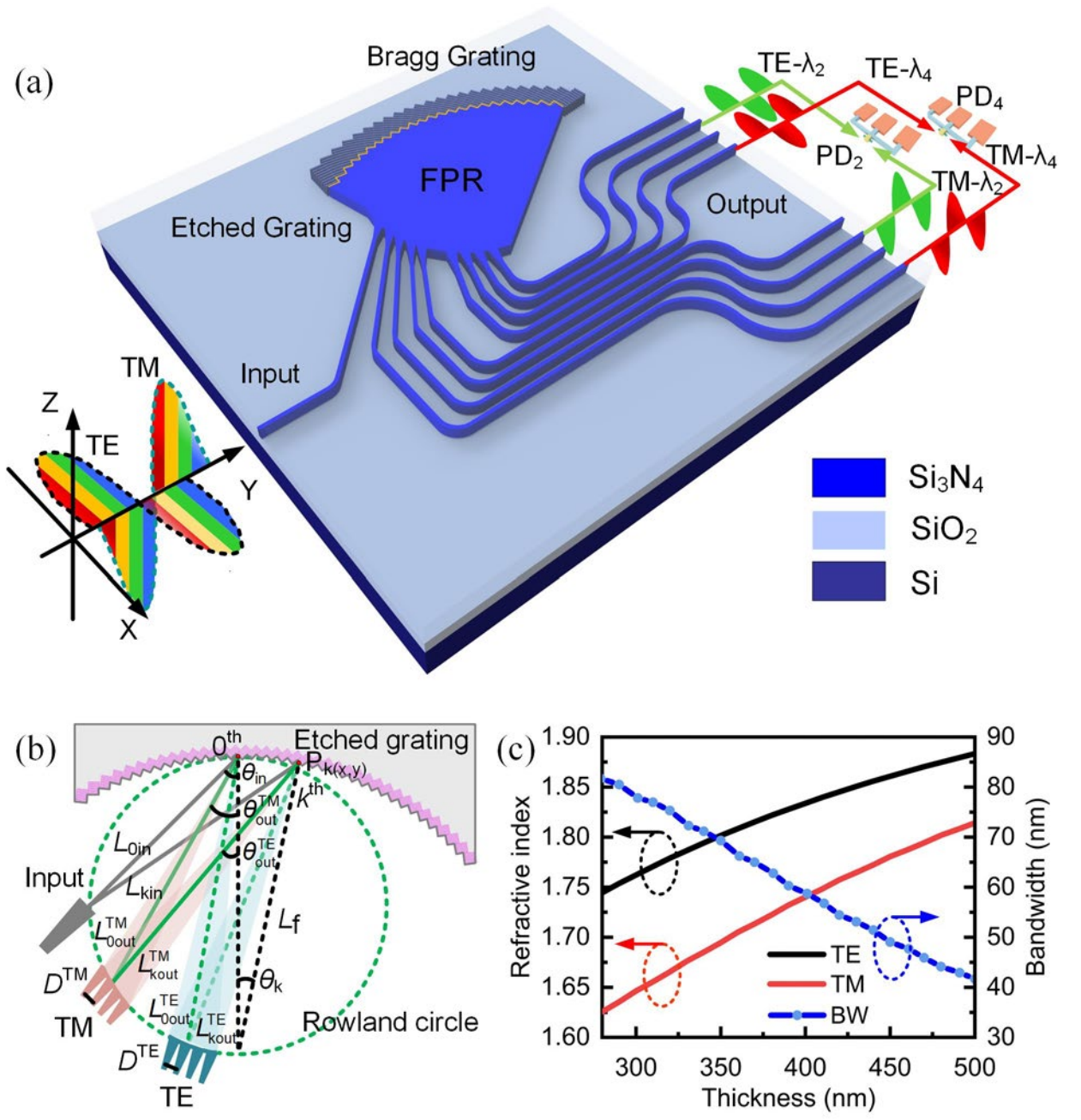

**Fig.1.** Device architecture and design principle. (a) Schematic of polarization-independent EDG demultiplexer. When multiplexed multi-wavelength light beams in arbitrary SOPs are input, they are demultiplexed into light beams with single TE/TM polarization and single wavelength and output on different channels. (b) Schematic of polarization and wavelength separation. (c) Effective refractive indices of TE/TM polarizations and working bandwidth of polarization separation corresponding to the thicknesses of the $Si_3N_4$ layer at 1310 nm wavelength.

The design of an EDG is commonly based on the principles of blazed grating and Roland mounting [32,33]. Our EDG is first identified with TE polarization and its operation is then analyzed in TM polarization, as shown in Fig. 1(b). The light output angle $\theta_{out}$ satisfies the following relationship with the effective refractive index $n_{eff}$ of the FPR slab waveguide, which is polarization-dependent.

$$(L_{kin} + L_{f} \cdot \sqrt{(\sin\theta_k + \cos(\theta_{out}) \cdot \sin(\theta_{out}))^2 + (\cos\theta_k + \sin(\theta_{out})^2)^2}) - (L_{0in} + L_{f} \cdot \cos(\theta_{out})) = \frac{km\lambda}{n_{eff}} \quad (1)$$

In the above equation $L_{kin}$, $L_{0in}$, $\theta_k$, $k$, $m$, $\lambda$, and $L_f$ are definite and independent of the polarization (see Appendix A). Thus, the light output angle $\theta_{out}$ can be determined by the effective refractive index $n_{eff}$ of the FPR slab waveguide. Owing to the large difference in $n_{eff}$ between the TE and TM polarizations, $\theta_{out}$ of the TE- and TM-polarized light beams are different, which indicates that TE and TM light beams as well as lights of different wavelengths are output at different positions of the Roland circle, that is, light signals of different polarizations and wavelengths can be completely separated.

Further design is needed for separating the polarizations in the desired wavelength range. The polarization dispersion in the FPR slab waveguide can be regulated by changing its material and thickness. In this study, $Si_3N_4$ was selected because of its low propagation loss, low thermo-optical effect, and CMOS compatibility [28,34]. The effective refractive indices of the TE and TM fundamental modes in the FPR slab waveguides of different $Si_3N_4$ layer thicknesses were simulated, as shown in Fig. 1(c). As the $Si_3N_4$ layer becomes thinner, the refractive index difference between the TE and TM polarizations significantly increases, resulting in a larger separation between output angles under the TE and TM polarizations according to equation (1).

In this case, the working bandwidth in the O-band, which is defined as the range of wavelength when the outputs of the two polarizations do not overlap, also increases a lot, as shown in the calculated results in Fig. 1(c). However, to achieve lower loss and crosstalk of the EDG demultiplexer, one-dimensional Bragg gratings behind the etched grating teeth with a thicker $Si_3N_4$ layer were required to enhance their reflectivity. Thus, a balanced $Si_3N_4$ layer thickness was obtained as 310 nm (details can be found in Appendix B). Moreover, the input and output waveguides were connected to the FPR with the tapered waveguides to reduce coupling loss. Several geometric parameters were carefully optimized, including the Bragg grating period (418 nm), diffraction order ($m$=3), and output waveguide spacing ($d$ = 5 μm) (details can be found in Appendix B). Finally, a low-loss O-band CWDM demultiplexer was obtained for polarization-independent wavelength demultiplexing.

## 3. SIMULATION RESULTS

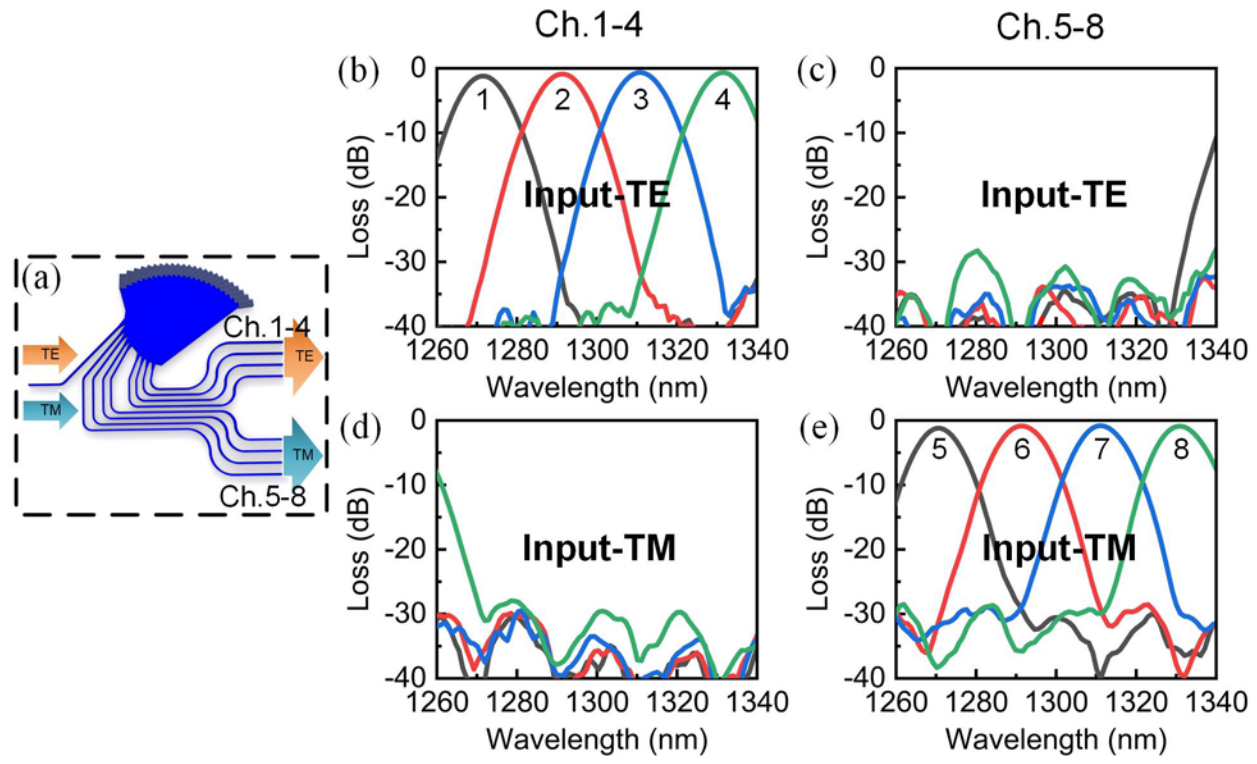

**Fig.2.** Simulated transmission spectrum of polarization-independent demultiplexing EDG. (a) Schematic of EDG model. (b) Output from channels 1–4 with TE input. (c) Output from channels 5–8 with TE input. (d) Output from channels 1–4 with TM input. (e) Output from channels 5–8 with TM input.

The operation of the polarization-independent EDG demultiplexer was verified through simulations with the Lumerical 2.5D finite-difference time-domain software, as shown in Fig. 2. When the multi-wavelength multiplexed light with TE polarization was input, the demultiplexed light beams were output from channels 1–4, whereas channels 5–8 generated very less output. On the contrary, in the case of TM polarization, light beams were output from channels 5–8 whereas limited output was obtained from channels 1–4. This fact shows that the TE/TM polarizations were demultiplexed along with the wavelengths in the EDG. The insertion losses of the channels were estimated to be 1.2 dB (1271.1 nm), 0.9 dB (1291.7 nm), 0.6 dB (1311.1 nm), and 0.7 dB (1331.0 nm) for the TE-polarized signals, whereas the values for the TM-polarized signals were 1.2 dB (1271.1 nm), 0.9 dB (1291.7 nm), 0.8 dB (1311.1 nm), and 0.9 dB (1331.0 nm). The device demonstrated good loss uniformity of approximately 0.5 dB and low crosstalk (better than −28 dB). The PDL was less than 0.2 dB, which indicates excellent polarization-independent performance. It should be noted that owing to the TE and TM polarizations being separated into different channels, the output positions for the TE and TM polarizations can be adjusted independently according to the linear dispersion on the Roland circle. Thus, a larger degree of design freedom can be obtained, and the dependence of the EDG wavelength shift on the polarization can be easily reduced to nearly zero.

## 4. EXPERIMENTAL METHODS

The proposed demultiplexer was fabricated on a silicon nitride platform with a $Si_3N_4$ core layer thickness of 310 nm and a buried oxide layer thickness of 2 μm. The $Si_3N_4$ layer was deposited with $SiH_4$ and $N_2O$ through plasma-enhanced chemical vapor deposition. Waveguides with a width of 800 nm and EDG patterns were defined using electron-beam lithography followed by full etching of the $Si_3N_4$ layer through inductively coupled plasma (ICP) dry etching. An 1-μm-thick $SiO_2$ layer was deposited on the device as the top cladding layer. The footprint of the fabricated EDG is 320 × 230 $\mu m^2$, and the scanning electron microscope (SEM) images of the device are shown in Fig. 3. The left and right images are magnified views of the rectangular areas in the center image. The left image depicts the input and output waveguides connected to the FPR with tapered waveguides, while the right image depicts the Bragg gratings behind the etched grating teeth.

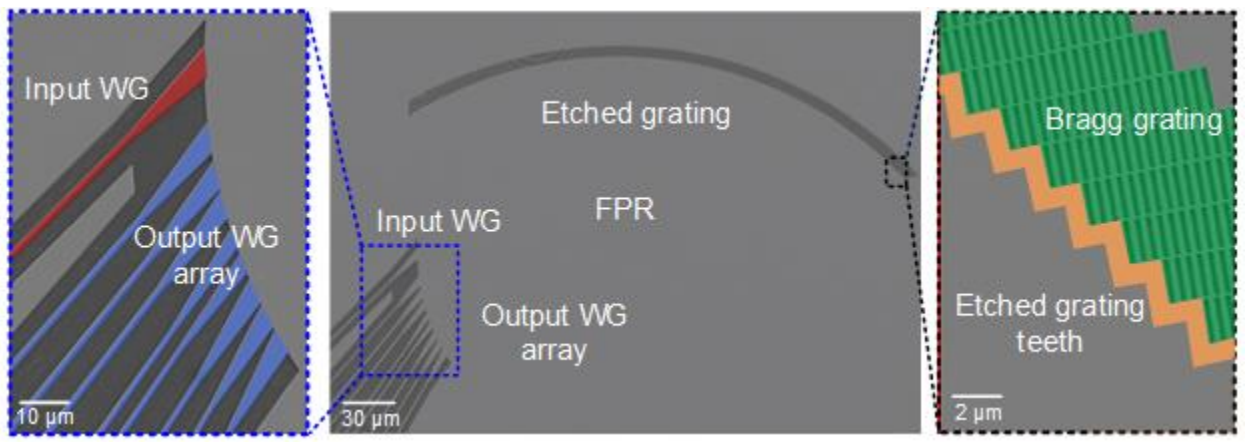

**Fig.3.** Scanning electron microscope image of EDG after ICP etching. Inset: SEM images of waveguide array attached to the FPR and boundary of etched grating teeth.

A wavelength-tunable laser (Santec, TSL-510), optical power meter (Yokogawa, AQ2211), polarization synthesizer (General Photonics, PSY-201), and end-face fiber coupling test systems were used for the measurements. The experimental setup is shown in Fig. 4(a). In the measurements, the wavelength-tunable laser outputs light of the target wavelength to the polarization synthesizer, which controls the polarization of the light and outputs light with the target SOP to the end-face fiber coupling system, which couples the light with the EDG chip. The signals at the output end of the chip are coupled with the receiver fiber and transmitted to the optical power meter. All instrument control and data recording were realized through a computer. Reference waveguides were added to the chip for system loss measurement to normalize the EDG device losses. The reference waveguide consisted of only straight and bent waveguides, where the bending radius and length of the bent waveguide were the same as those for the EDG. With the reference waveguides, the fiber–waveguide coupling loss, straight waveguide loss, and bent waveguide loss can be filtered to obtain the normalized on-chip insertion loss of the EDG device.

## 5. EXPERIMENTAL RESULTS AND DISCUSSION

### A. Device Testing in TE/TM Polarization

The EDG demultiplexer was characterized using the experimental setup shown in Fig. 4(a). The normalized transmission spectra in Fig. 4(c)–(f) show that the EDG successfully separated the TE and TM multiplexed multi-wavelength lights into different output channels, which agree with the simulation results shown in Fig. 2. The peak output light wavelengths were 1284.2 nm, 1303.9 nm, 1322.9 nm, and 1342.3 nm for the TE-polarized input and 1283.2 nm, 1303.5 nm, 1322 nm, 1341 nm for the TM-polarized input. Compared with the simulations, around 12 nm red shifts of the output wavelengths were observed, which were considered to arise from the unexpected change in thickness of the $Si_3N_4$ layer during the $Si_3N_4$ film deposition. According to the error analysis (shown in Appendix C), a change in thickness of approximately 10 nm caused a shift of 6–8 nm in the wavelength in the EDG output due to a change in the effective refractive index. And the measured thickness of the $Si_3N_4$ layer showed a variation of approximately 19.2 nm in this study, which agreed well with the simulation. Nevertheless, the EDG wavelength intervals remained close to the designed value of 20 nm. The insertion losses of the channels were 2.4 dB, 2.3 dB, 2.3 dB, and 1.3 dB for the TE-polarized input, and 1.9 dB, 1.1 dB, 0.5 dB, and 2.2 dB for the TM-polarized input, with PDLs of 0.5–1.8 dB. The crosstalks were better than −30 dB for both TE and TM polarizations. Compared with the simulations, the increased insertion losses and PDLs mainly resulted from the perpendicularity and roughness of the sidewalls of the Bragg gratings, which can be improved by further optimization of the fabrication process.

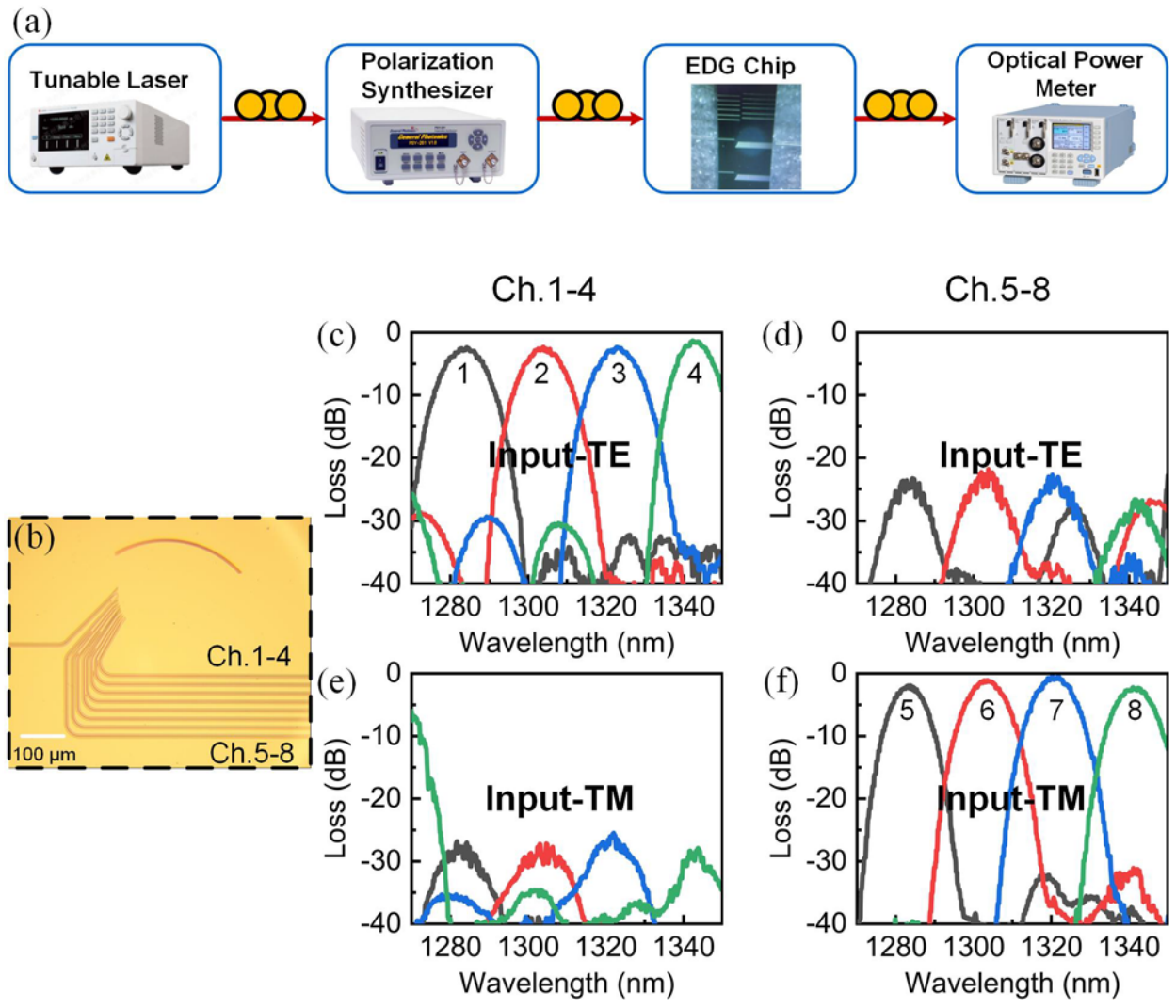

**Fig.4.** EDG measurements. (a) Experimental setup. (b) Optical microscope view of the fabricated EDG. (c)–(f) Transmission spectra. (c) Outputs from channels 1–4 with TE input. (d) Outputs from channels 5–8 with TE input. (e) Outputs from channels 1–4 with TM input. (f) Outputs from channels 5–8 with TM input.

## B. Device Testing in Arbitrary SOPs

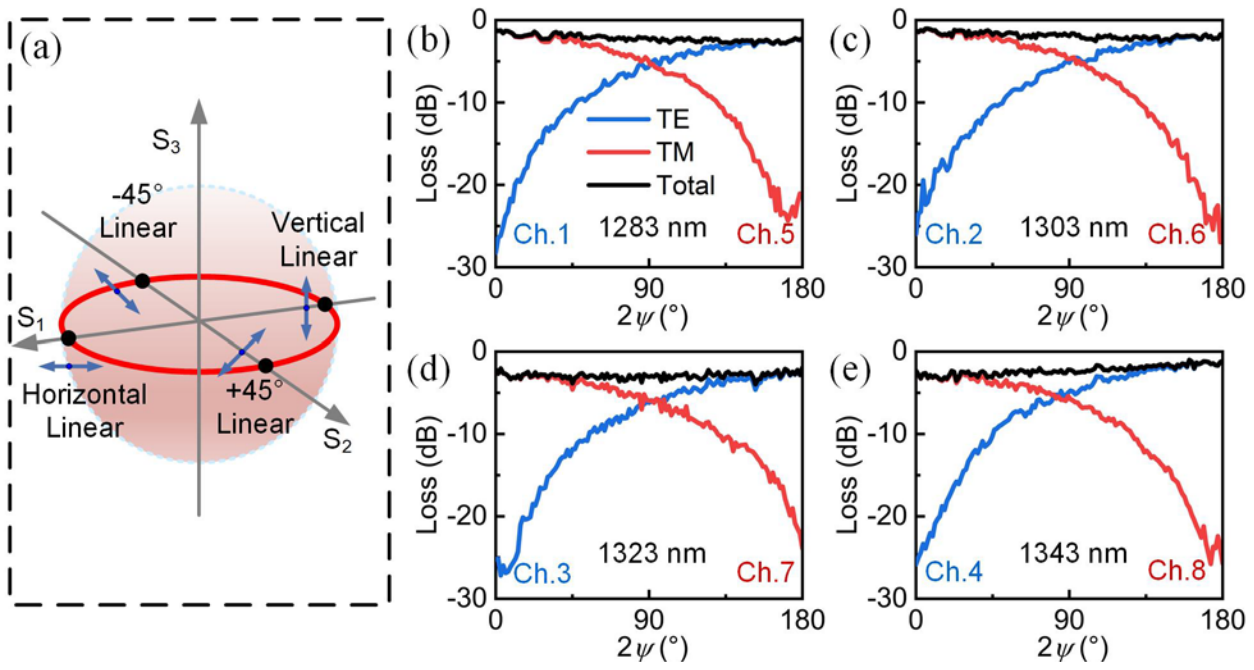

**Fig.5.** Scanning measurement along the RHC trajectory (red line) with azimuthal angles $2\psi \in (0, \pi)$ on the horizontal axis and losses on the vertical axis. The outputs (black solid) are superimposed by the TE (blue solid) and TM (red solid) outputs of the same wavelength. (a) Polarization state represented by Poincaré sphere. (b)–(e) Transmission spectra at output wavelengths. (b) Channels 1 and 5 at 1283 nm. (c) Channels 2 and 6 at 1303 nm. (d) Channels 3 and 7 at 1323 nm. (e) Channels 4 and 8 at 1343 nm.

To test the performance of the EDG device in arbitrary SOPs, the SOP of the input light was scanned with a polarization synthesizer, which could generate the desired SOP signals at fixed wavelengths. The receding horizontal control (RHC) trajectory (see Fig. 5(a)) was scanned so that the SOP of the input light evolved from TM linear polarization to TE linear polarization, with the azimuth angle changing gradually from 0° to 180°. The results are shown in Fig. 5(b)–(e). Despite the TE and TM output intensities showing large changes with opposite trends according to Malus's law, the total outputs of the two polarization channels at the same wavelength (black solid) were close to constant values of 2.21 dB, 1.85 dB, 2.88 dB, and 2.27 dB with standard deviations of 0.39 dB, 0.35 dB, 0.37 dB, and 0.54 dB, respectively, for the four wavelengths. This indicates that the optical signal that the PD received from the superimposed light remained stable at arbitrary linearly polarized input lights.

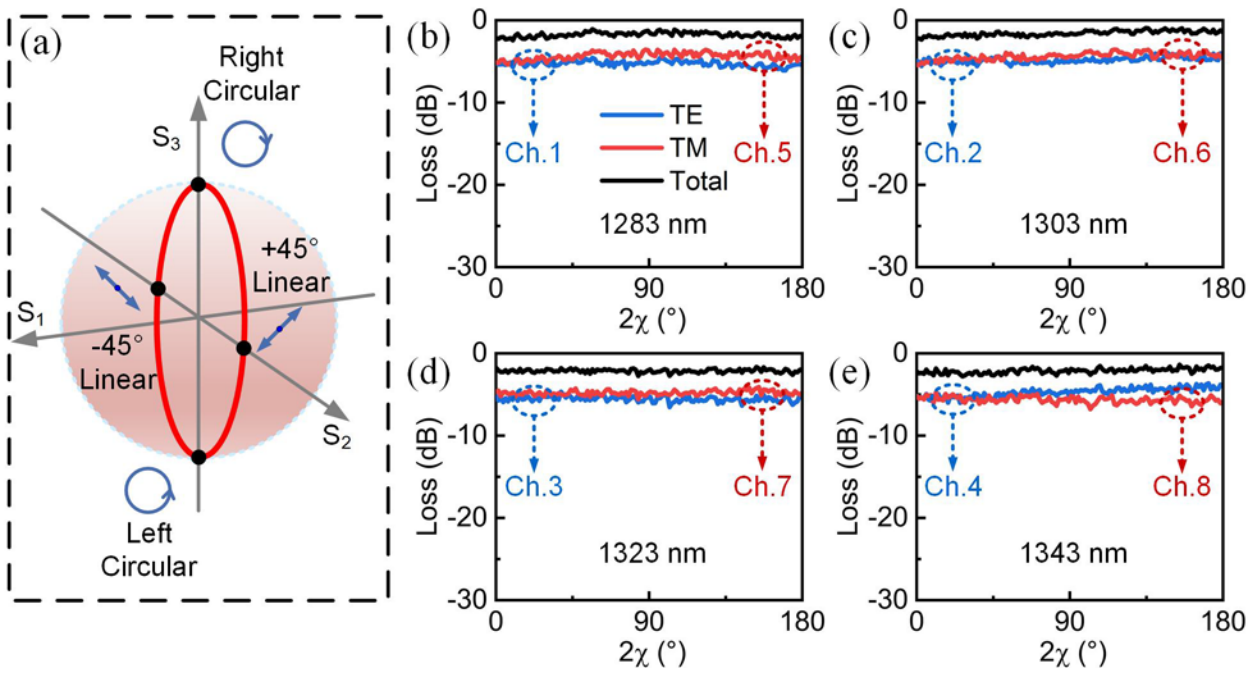

**Fig.6.** Scanning measurement along the LP0 trajectory (red line) with ellipticity (0, π) on the horizontal axis and device loss on the vertical axis. (a) Polarization state represented by Poincaré sphere. (b)-(e) Transmission spectrum. (b) Outputs of channels 1 and 5 at 1283 nm. (c) Outputs of channels 2 and 6 at 1303 nm. (d) Outputs of channels 3 and 7 at 1323 nm. (e) Outputs of channels 4 and 8 at 1343 nm.

Moreover, the influence of the polarization phase on the device was investigated, as shown in Fig. 6. The polarization phase of the input light was changed so that the input SOP changed from −45° linear polarization through elliptical polarization and circular polarization and then to 45° linear polarization along the LP0 trajectory, whereas the intensities of the TE and TM polarizations were fixed and equal. Fig. 6 shows that the superimposed outputs of the TE and TM polarizations remained stable at 1.75 dB, 1.58 dB, 2.13 dB, and 2.12 dB for the four wavelengths with variation in the polarization phase, with standard deviations of 0.29 dB, 0.31 dB, 0.20 dB, and 0.30 dB, respectively. This shows that the change in the polarization phase has a weak effect on the device's operation. Hence, the EDG can operate independently with any linearly, elliptically, and circularly polarized light input. Based on the experiments shown above, a polarization-independent wavelength demultiplexer with high efficiency was successfully demonstrated with a single EDG device.

## 6. CONCLUSION

This paper proposed a novel design of polarization-independent wavelength demultiplexer by utilizing the birefringence effect, which can be extended to numerous planar wavelength demultiplexing devices such as EDGs and arrayed waveguide gratings. The design principle was successfully applied on an EDG device on a silicon-based $Si_3N_4$ platform, and a polarization-independent wavelength EDG demultiplexer was experimentally fabricated and demonstrated. Measurements by scanning the polarizations of input lights showed that stable outputs could be obtained at four demultiplexed wavelengths with standard deviations of 0.4–0.5 dB for the equator of the Poincaré sphere and 0.2–0.3 dB for the longitude of the Poincaré sphere, which demonstrated the good polarization-independent property of the EDG device. The insertion losses of the EDG device were measured to be 0.5–2.4 dB with PDLs of 0.5–1.8 dB, and the crosstalk was better than −30 dB at the four output wavelengths. This compact and low-loss polarization-independent wavelength demultiplexer, which can be monolithically integrated with PDs and other devices on a SiPh platform, has broad applicability in telecommunication and data communication systems. The design scheme of using the birefringence effect, proposed by us, not only significantly simplifies the device structure of the demultiplexers in WDM receivers but also provides a promising solution for designing other polarization-independent devices on the SiPh platform.

## APPENDIX A: DESIGN DETAILS OF SIMULTANEOUS DEMULTIPLEXING OF POLARIZATION AND WAVELENGTH

The design of the EDG demultiplexer is based on the principles of blazed grating and Rowland mounting [32, 33]. When the difference in the effective optical path is designed as an integer multiple of the wavelength, the diffracted light will be enhanced and light of different wavelengths can be demultiplexed in the EDG. The working principle of a single-polarization EDG is shown in Fig. 7(a), and the grating teeth located at the tangential points of the Rowland circle (green dotted line) and grating circle (black dotted line) are considered as the central grating teeth ($0_{th}$). In the case of single-polarization EDGs (such as TE polarization), the following equations need to be satisfied.

$$(L_{\text{kin}} + L_{\text{kout}}^{\text{TE}}) - (L_{0\text{in}} + L_{0\text{out}}^{\text{TE}}) = \frac{km\lambda}{n_{\text{eff}}^{\text{TE}}}, \tag{A1}$$

$$L_{0\text{in}} = L_{\text{f}} \cdot \cos(\theta_{\text{in}}), \tag{A2}$$

$$L_{0\text{out}}^{\text{TE}} = L_{\text{f}} \cdot \cos(\theta_{\text{out}}^{\text{TE}}), \tag{A3}$$

$$L_{\text{kin}} = L_{\text{f}} \cdot \sqrt{(\sin(\theta_{\text{k}}) + \cos(\theta_{\text{in}}) \cdot \sin(\theta_{\text{in}}))^2 + (\cos(\theta_{\text{k}}) + \sin(\theta_{\text{in}})^2)^2}, \tag{A4}$$

$$L_{\text{kout}}^{\text{TE}} = L_{\text{f}} \cdot \sqrt{(\sin(\theta_{\text{k}}) + \cos(\theta_{\text{out}}^{\text{TE}}) \cdot \sin(\theta_{\text{out}}^{\text{TE}}))^2 + (\cos(\theta_{\text{k}}) + \sin(\theta_{\text{out}}^{\text{TE}})^2)^2}, \tag{A5}$$

where $L_{\text{f}}$ is the diameter of the Rowland circle [35,36], $L_{0\text{in}}$ and $L_{0\text{out}}$ are the effective incident and output optical paths of the central grating tooth, $L_{\text{kin}}$ and $L_{\text{kout}}$ are the effective incident and output optical paths of the $k_{\text{th}}$ grating tooth, and $\theta_{\text{in}}$ and $\theta_{\text{out}}$ are the angle of incidence and diffraction, respectively. $\theta_{\text{k}}$ is the rounding angle of grating circle corresponding to the midpoint of each grating tooth, $m$ is the diffraction order, $\lambda$ is the wavelength of the incident light, and $n_{\text{eff}}$ is the effective index of the propagation medium, which is usually considered as the effective refractive index of the fundamental mode of the slab waveguide. Therefore, $\theta_{\text{k}}$ can be obtained using Equation (A1), and the arrangement of the etched grating can be determined as shown in existing literature.

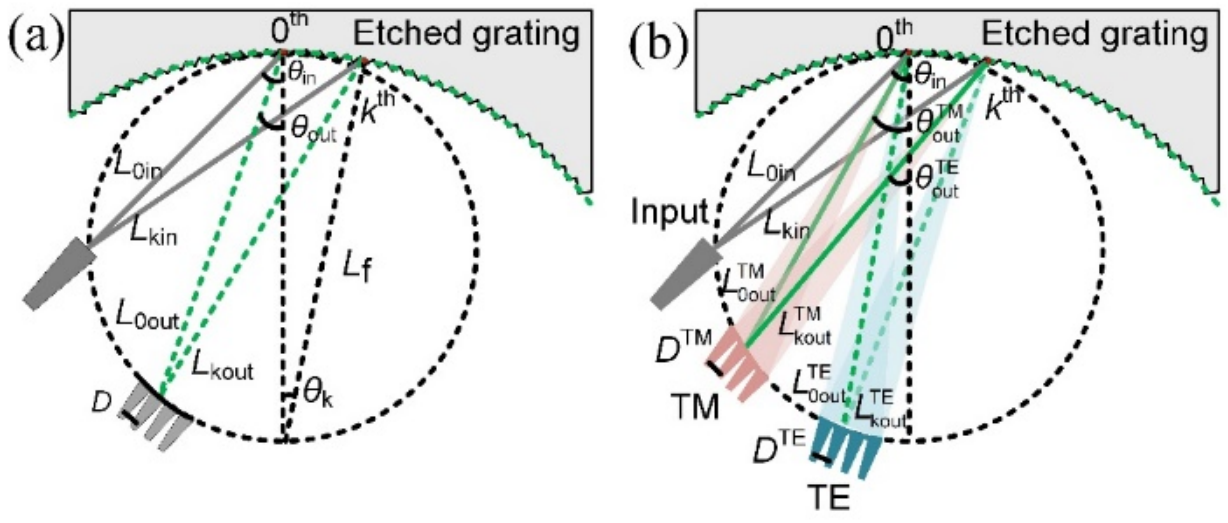

**Fig. 7.** Schematic of EDG design rules. (a) EDG design under single polarization. (b) EDG design under TE and TM polarizations.

Further, we utilize the polarization dispersion derived from birefringence to design the polarization-independent EDG, as shown in Fig. 7(b). In the case of TM polarization, a similar phase difference relationship can be considered to realize wavelength demultiplexing.

$$(L_{\text{kin}} + L_{\text{kout}}^{\text{TM}}) - (L_{0\text{in}} + L_{0\text{out}}^{\text{TM}}) = \frac{km\lambda}{n_{\text{eff}}^{\text{TM}}}, \tag{A6}$$

$$L_{0\text{out}}^{\text{TM}} = L_{\text{f}} \cdot \cos(\theta_{\text{out}}^{\text{TM}}), \tag{A7}$$

$$L_{\text{kout}}^{\text{TM}} = L_{\text{f}} \cdot \sqrt{(\sin(\theta_{\text{k}}) + \cos(\theta_{\text{out}}^{\text{TM}}) \cdot \sin(\theta_{\text{out}}^{\text{TM}}))^2 + (\cos(\theta_{\text{k}}) + \sin(\theta_{\text{out}}^{\text{TM}})^2)^2}, \tag{A8}$$

where $L_{0out}^{TM}$ and $L_{kout}^{TM}$ are the output optical paths of the central grating tooth and $k_{\text{th}}$ grating tooth, respectively, in TM polarization. Therefore, the output angle in TM polarization, $\theta_{\text{out}}^{\text{TM}}$, can be calculated by substituting the positions of the grating teeth and effective refractive index of the slab waveguide in TM polarization, and the same diffraction order can be selected for uniform diffractions. Comparing Equations (A1) and (A6), the output angle at different polarizations is dependent only on their refractive indices. Thus, an EDG with simultaneous demultiplexing of polarization and wavelength can be realized.

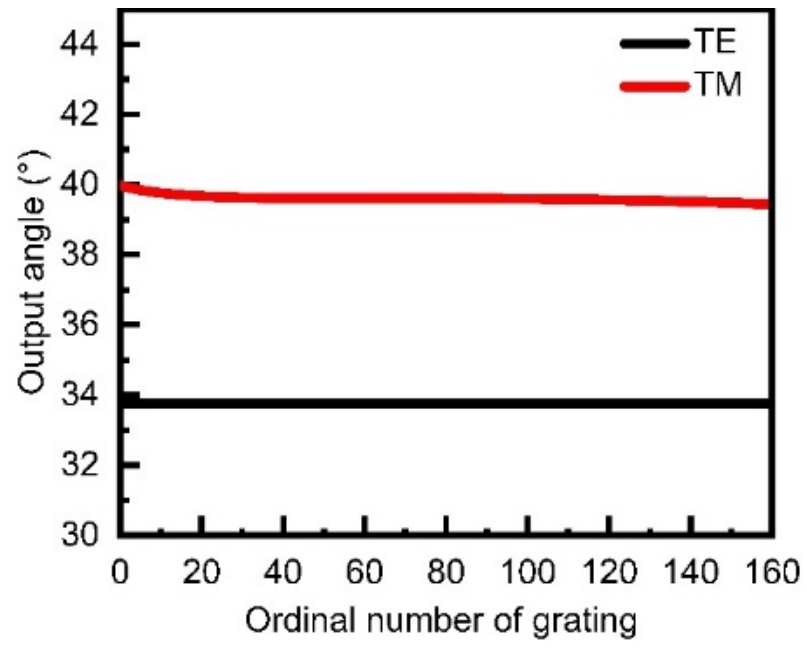

**Fig. 8.** Aberration analysis under TE and TM polarizations.

However, according to Equations (A6)–(A8), selecting different angles $\theta_k$ of the grating teeth may change $\theta_{out}^{TM}$, which causes aberrations and increased loss. Therefore, the aberrations of the output under TM polarization should be considered. We calculated the output angles corresponding to each grating tooth of the EDG in both TE and TM polarizations, as shown in Fig. 8. We can see that the output angles of the grating teeth remain the same in TE polarization. In TM polarization, there are slight changes in the output angles of the grating teeth at the edges, whereas the output angles are almost constant at the central grating teeth, where the light emissions are much stronger. Therefore, we define the output angles according to the positions of the central grating teeth to obtain minimal aberration and better device performance.

## APPENDIX B: HIGH-PERFORMANCE EDG DESIGN

### 1. Appropriate $Si_3N_4$ Thickness Design for Multidimensional Multiplexing

The thickness of the $Si_3N_4$ layer has a significant influence on the performance of the Bragg gratings and polarization dispersion. The influence of thickness on the performance of the Bragg grating was simulated, as shown in Fig. 9(a). With the thickening of the $Si_3N_4$ layer, the reflection bandwidth of the Bragg gratings increases. To reduce the insertion loss, the reflection bandwidth of the Bragg gratings should cover the wavelength range for demultiplexing in both TE and TM polarizations, which requires that the $Si_3N_4$ thickness $t_{Si3N4}$ should be larger than 250 nm.

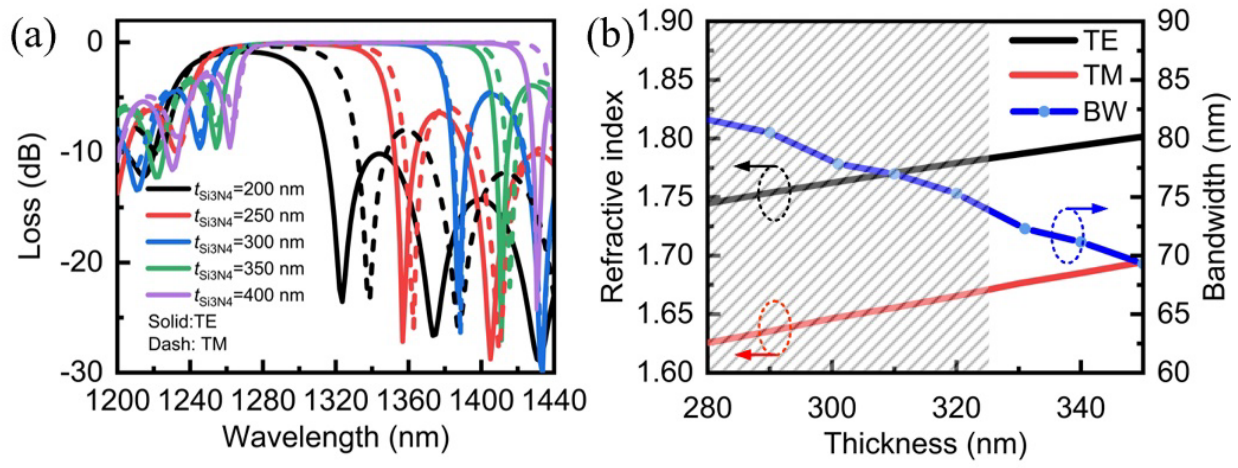

**Fig. 9.** (a) Reflection of Bragg gratings at different $Si_3N_4$ thicknesses (period of Bragg grating is 418 nm). (b) Effective refractive indices of TE/TM polarizations and working bandwidth of polarization separation corresponding to the thickness of the $Si_3N_4$ layer at 1310 nm wavelength. The gray area is the thickness for CWDM4 demultiplexing.

The working bandwidths are shown in Fig. 9(b). When the $Si_3N_4$ layer becomes thinner, the difference in $n_{eff}$ between the TE and TM polarizations increases significantly, allowing polarization separation of the EDG outputs over a wider wavelength range. According to the provisions of ITU.G.694.2, $t_{Si3N4}$ should not exceed 325 nm to obtain sufficient separation bandwidth for CWDM4 demultiplexing in O-band. Therefore, we finally selected $t_{Si3N4}$ = 310 nm by weighing the reflection efficiency of the Bragg gratings and working bandwidth of the EDG.

## 2. Diffraction Order Design for Low Insertion Loss and PDL

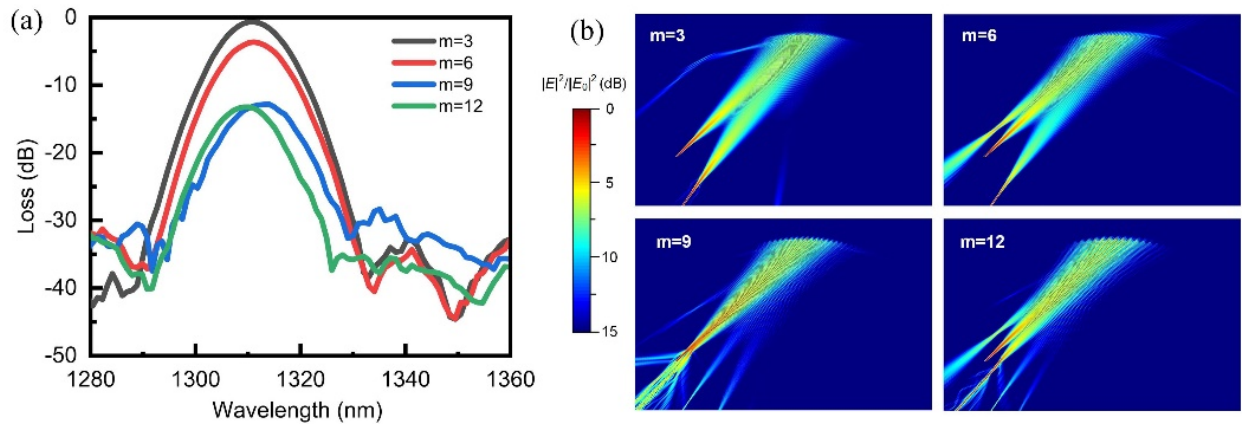

**Fig. 10.** (a) Output spectra of the central output channel under TE polarization for different diffraction orders. (b) Electric field distribution at $\lambda$=1311 nm under TE polarization for different diffraction orders.

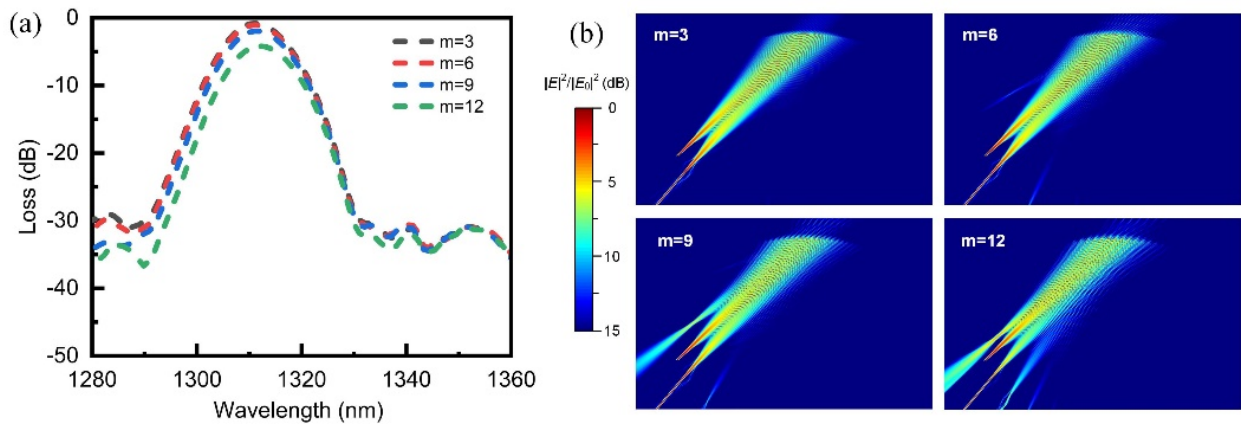

**Fig. 11.** (a) Output spectra of the central output channel under TM polarization for different diffraction orders. (b) Electric field distributions at $\lambda$=1311 nm under TM polarization for different diffraction orders.

As shown in Fig. 10 and Fig. 11, the performances under different diffraction orders were estimated to optimize the diffraction order $m$ of the EDG. When $m$ gradually increases, the number of grating teeth as well as the free spectral range decreases and causes multiple diffraction orders, which are close to the designed output position. Therefore, when $m$ is large, adjacent diffraction orders will absorb the energy of the diffraction order designed by us, resulting in increased loss. As shown in Fig. 10(b) and Fig. 11(b), the EDG exhibits ultra-low insertion losses in both polarizations at $m$=3. When $m$ is larger, the insertion loss of the EDG increases significantly, especially under TE polarization. Moreover, a smaller diffraction order will result in grating teeth that are more compact, which complicates the fabrication process without significant performance improvements. Considering these facts, we selected $m$=3.

## 3. Output Waveguide Spacing Design for Low Crosstalk

Crosstalk is defined as the ratio of the output power at the central wavelength of the channel to the output power at this wavelength of the other channels. As shown in Fig. 12, to optimize the crosstalk of the EDG, we simulated the output spectra under TE and TM polarizations at 1291 nm (solid line) and 1311 nm (dashed line), respectively, for different values of the output waveguide spacing $d$. As $d$ increases, the crosstalk of the device under each polarization shows a decreasing trend. In addition, we noticed that the noise floor under TM polarization is higher than that under TE polarization. To obtain the ideal crosstalk characteristics in both polarizations as well as a compact footprint, $d$=5 μm was finally selected as the optimal parameter.

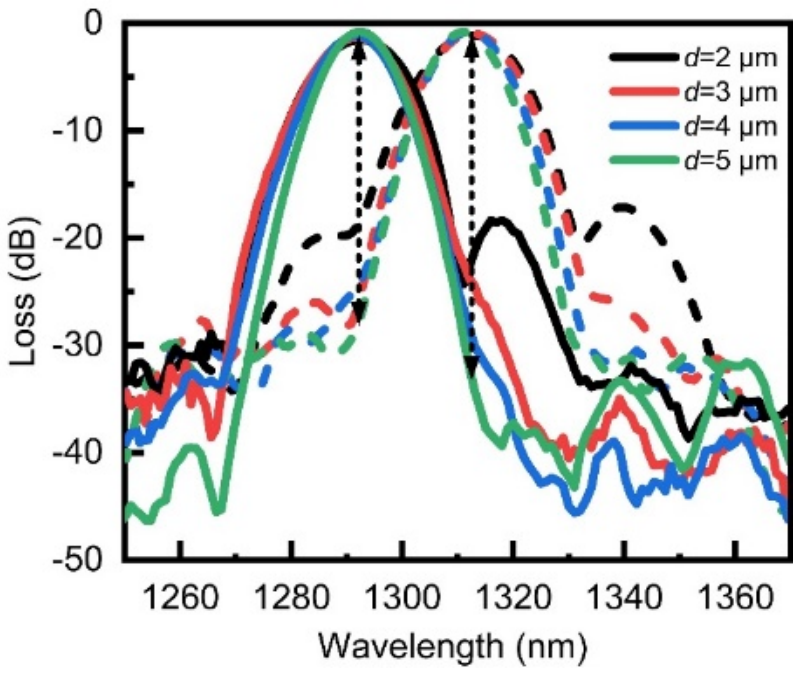

**Fig. 12.** Simulated results of the output spectra of TE (solid line) and TM (dashed line) polarizations at different values of the output waveguide spacing.

## APPENDIX C: EXPERIMENTAL ERROR ANALYSIS

During fabrication, the unexpected variation in the thickness of the $Si_3N_4$ layer during film deposition can significantly affect the output wavelengths of the EDG. Therefore, we analyzed the variations in the TE/TM output spectra of the central output channels for different $Si_3N_4$ thicknesses, as shown in Fig. 13 According to the simulations, an increase of 10 nm in the thickness causes overall red shifts of approximately 6–8 nm of the EDG output spectra in both TE and TM polarizations.

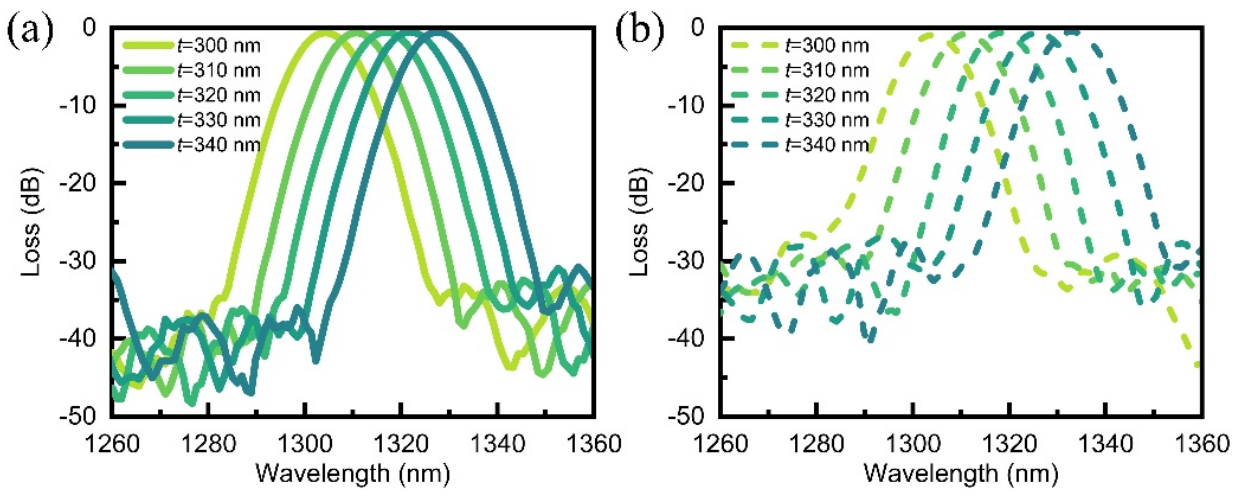

**Fig. 13.** Simulated results of the output spectra of TE (solid line) and TM (dashed line) polarizations for central channel for various $Si_3N_4$ layer thicknesses.

During the actual $Si_3N_4$ deposition process, the thicknesses of the $Si_3N_4$ layer grown in 870 s were 306.5 nm and 325.7 nm owing to the limitation related to the accuracy of the plasma-enhanced chemical vapor deposition equipment. The difference in thickness of 19.2 nm caused a shift in the output wavelengths of the fabricated EDGs. Therefore, the consistency of the output wavelength of the fabricated device can be further improved by using deposition equipment of greater accuracy.

## APPENDIX D: COMPARISON WITH RECENT DEMULTIPLEXERS

The comparison of the device proposed in this paper with other polarization-independent demultiplexing devices reported in recent years is shown in Table 1. From this, we can see that our device was superior in terms of both performance parameters such as insertion loss, crosstalk, and PDL, as well as large-scale application requirements such as process compatibility and footprints. Therefore, it is a promising solution for the implementation of SiPh WDM receivers.

**TABLE 1. Comparison of Polarization-Independent Demultiplexing Devices**

| Ref. | Component | CMOS compatible | Cascaded or not | Insertion loss (dB) | PDL (dB) | Crosstalk (dB) | Channel number /spacing | Footprint ($mm^2$) |
|---|---|---|---|---|---|---|---|---|
| **This work** | **EDG** | **Yes** | **No** | **0.5** | **0.5~1.8** | **-30** | **4/20 nm** | **0.07** |
| [21] | PBS+PR+AWG | Yes | Yes | 6.5 | 2.0 | -12 | 8/0.8 nm | 20 |
| [37] | PSR+AWG | Yes | Yes | 3.6 | 1.2~1.8 | -15 | 8/10 nm | N/A |
| [24] | PBS+PR+ DMZI+AWG | Yes | Yes | 5.5 | 0.5 | -20 | 16/0.8 nm | 4.76 |
| [25] | PBS+AWG | Yes | Yes | 6 | 0.2 | -16 | 8/0.8 nm | ~6.5 |
| [28] | PR+AWG | Yes | Yes | 2.2 | 0.2~0.7 | -29 | 4/20 nm | 1.03 |
| [29] | EDG | Yes | No | 2 | 0.2~0.7 | -25 | 40/0.8 nm | 150 |
| [8] | Silica AWG | No | No | 2.5 | N/A | -30 | 4/20 nm | 29.4 |
| [9] | Silica AWG | No | No | 2 | 0.2 | -20 | 8/0.8 nm | ~50 |

## Funding

National Ministry of Science and Technology Key Research and Development Program (2018YFB2201200); National Natural Science Foundation of China (61635011).

## Acknowledgements

The authors acknowledge Zhejiang University Micro and Nano Processing Platform for providing the facility support. The authors thank Dr. Wei Ma, Dr. Tong Ye, Mr. Lin Han, and Mrs. Ying Huang for their fruitful discussions on the experiments conducted in this study and revisions made in the manuscript.

## Disclosures

The authors declare no conflicts of interest.

## Data Availability

The data that support the findings of this study are available from the corresponding author upon reasonable request.

## References

1. E. Kuramochi, K. Nozaki, A. Shinya, K. Takeda, T. Sato, S. Matsuo, H. Taniyama, H. Sumikura, M. Notomi, “Large-scale integration of wavelength-addressable all-optical memories on a photonic crystal chip,” Nature Photonics **8**, 474–481 (2014).
2. D. Dai, J. E. Bowers, “Silicon-based on-chip multiplexing technologies and devices for Peta-bit optical interconnects,” Nanophotonics **3**, 283–311 (2014).
3. J. M. Kahn, K. P. Ho, “A bottleneck for optical fibres,” Nature **411**, 1007–1010 (2001).
4. R. Nagarajan, C. H. Joyner, R. P. Schneider, J. S. Bostak, T. Butrie, A. G. Dentai, V. G. Dominic, P. W. Evans, M. Kato, M. Kauffman, D. J. H. Lambert, S. K. Mathis, A. Mathur, R. H. Miles, M. L. Mitchell, M. J. Missey, S. Murthy, A. C. Nilsson, F. H. Peters, S. C. Pennypacker, J. L. Pleumeekers, R. A. Salvatore, R. K. Schlenker, R. B. Taylor, T. Huan-Shang, M. F. Van Leeuwen, J. Webjorn, M. Ziari, D. Perkins, J. Singh, S. G. Grubb, M. S. Reffle, D. G. Mehuys, F. A. Kish, D. F. Welch, “Large-scale photonic integrated circuits,” IEEE J. Sel. Top. Quantum Electron. **11**, 50–65 (2005).
5. B. Huttner, C. Geiser, N. Gisin, “Polarization-induced distortions in optical fiber networks with polarization-mode dispersion and polarization-dependent losses,” IEEE J. Sel. Top. Quantum Electron. **6**, 317 (2000).
6. C. Xie, L. F. Mollenauer, “Performance degradation induced by polarization-dependent loss in optical fiber transmission systems with and without polarization-mode dispersion,” J. Light. Technol. **21**, 1953 (2003).
7. C. H. Henry, R.F. Kazarinov, Y. Shani, R.C. Kistler, V. Pol, K.J. Orlowsky, “Four-channel wavelength division multiplexers and bandpass filters based on elliptical Bragg reflectors,” J. Light. Technol. **8**, 748 (1990).
8. L. Liu, L. Chang, Y. Kuang, Z. Li, Y. Liu, H. Guan, M. Tan, Y. Yu, Z. Li, “Low-cost hybrid integrated 4 × 25 GBaud PAM-4 CWDM ROSA with a PLC-based arrayed waveguide grating de-multiplexer,” Photonics Res. **7**, 722–727 (2019).
9. K. Ikeda, N. Matsubara, J. Hasegawa, R. Konoike, K. Suzuki, H. Kawashima, H. Matsuura, “5.5%-Δ-PLC/Silicon photonics hybrid wavelength MUX/DEMUX-and-switch device,” presented at Opt. Fiber Commun. Conf. (OFC), Washington, DC, June, 2021.

10. J. B. D. Soole, A. Scherer, H.P. LeBlanc, N.C. Andreadakis, R. Bhat, M.A. Koza, "Monolithic InP/InGaAsP/InP grating spectrometer for the 1.48–1.56 μm wavelength range," Appl. Phys. Lett. **58**, 1949–1951 (1991).
11. C. Cremer, G. Ebbinghaus, G. Heise, R. Müller-Nawrath, M. Schienle, L. Stoll, "Grating spectrograph in InGaAsP/lnP for dense wavelength division multiplexing" Appl. Phys. Lett. **59**, 627–629 (1991).
12. X. Chen, M. M. Milosevic, S. Stanković, S. Reynolds, T. D. Bucio, K. Li, "The emergence of silicon photonics as a flexible technology platform," Proc. IEEE **106**, 2101–2116 (2018).
13. C. Sciancalepore, R. J. Lycett, J. A. Dallery, S. Pauliac, K. Hassan, J. Harduin, H. Duprez, U. Weidenmueller, D. F. G. Gallagher, S. Menezo, B. B. Bakir, "Low-crosstalk fabrication-insensitive echelle grating demultiplexers on silicon-on-insulator," *IEEE Photon. Technol. Lett.* **27**, 494–497 (2015).
14. R. Cheng, C. L. Zou, X. Guo, S. Wang, X. Han, H. X. Tang, "Broadband on-chip single-photon spectrometer," Nat. Commun. **10**, 4104 (2019).
15. D. Melati, P. G. Verly, A. Delâge, S. Wang, J. Lapointe, P. Cheben, J. H. Schmid, S. Janz, D.X. Xu, "Compact and low crosstalk echelle grating demultiplexer on silicon-on-insulator technology." Electronics **8**, 687 (2019).
16. T. Ye, Y. Fu, L. Qiao, T. Chu, "Low-crosstalk Si arrayed waveguide grating with parabolic tapers," Opt. Express **22**, 31899–31906 (2014).
17. S. Cheung, M. R. Tan, "Ultra-low loss and fabrication tolerant silicon nitride ($Si_3N_4$) (de-)muxes for 1-μm CWDM optical interconnects," presented at Opt. Fiber Commun. Conf. (OFC), Diego, CA, U.S.A., March, 2020.
18. T. H. Yen, Y. Hung, "Fabrication-insensitive CWDM (de)multiplexer based on cascaded Mach-Zehnder interferometers," presented at Opt. Fiber Commun. Conf. (OFC), Diego, CA, U.S.A., March, 2020.
19. L. W. Luo, N. Ophir, C. P. Chen, L. H. Gabrielli, C. B. Poitras, K. Bergmen, M. Lipson, "WDM-compatible mode-division multiplexing on a silicon chip," *Nat. Commun*. **5,** 3069 (2014).
20. T. Barwicz, M. R. Watts, M. A. Popović, P. T. Rakich, L. Socci, F. X. Kärtner, E. P. Ippen, H. I. Smith, "Polarization-transparent microphotonic devices in the strong confinement limit," *Nat. Photonics* **1**, 57–60 (2007).
21. L. Chen, C. R. Doerr, Y. Chen, "Polarization-diversified DWDM receiver on silicon free of polarization-dependent wavelength shift," presented at Opt. Fiber Commun. Conf. (OFC), Los Angeles, CA, U.S.A., March, 2012.
22. L. Chen, "Silicon photonic integrated circuits for WDM technology and optical switch" presented at Opt. Fiber Commun. Conf. (OFC), Anaheim, CA, U.S.A., March, 2013.
23. S. H. Jeong, Y. Sobu, S. Tanaka, K. Morito, "WDM interconnect targeted Si-wire optical demultiplexers for large manufacturing tolerance, low voltage tunability and polarization diversified operability," presented at Opt. Fiber Commun. Conf. (OFC), Anaheim, CA, U.S.A., March, 2016.
24. S.H. Jeong, Y. Onawa, D. Shimura, H. Okayama, T. Aoki, H. Yaegashi, T. Horikawa, T. Nakamura, "Polarization diversified 16λ demultiplexer based on silicon wire delayed interferometers and arrayed waveguide gratings," *IEEE J. Light. Technol*. **38**, 2680–2687 (2020).
25. Q. Han, J. St-Yves, Y. Chen, M. Ménard, W. Shi, "Polarization-insensitive silicon nitride arrayed waveguide grating," Opt. Lett. **44,** 3976–3979 (2019).
26. K. Li, J. Zhu, Y. Mao, N. Zhang, X. Hou, "Design of efficient concave diffraction grating on 220 nm SOI platform for hybrid WDM–PDM (de)multiplexing." Opt. Commun. **477** 126358 (2020).
27. H. Xu, L. Liu, Y. Shi, "Polarization-insensitive four-channel coarse wavelength-division (de)multiplexer based on Mach–Zehnder interferometers with bent directional couplers and polarization rotators," Opt. Lett. **43,** 1483–1486 (2018).
28. S. Guerber, C. A. Alonso-Ramos, X. Le Roux, N. Vulliet, E. Cassan, "Polarization independent and temperature tolerant AWG based on a silicon nitride platform," Opt. Lett. **45**, 6559–6562 (2020).
29. D. Feng, Q. Wei, L. B. Hong, J. Luff, "High-speed receiver technology on the SOI platform," *IEEE J. Sel. Top. Quantum Electron*. **19**, 3800108–3800108 (2013).
30. N. Zhu, J. Song, L. Wosinski, S. He, "Design of a polarization-insensitive echelle grating demultiplexer based on silicon nanophotonic wires." *IEEE Photonics Technol. Lett*. **20**, 860–862 (2008).
31. J. Zou, X. Xia, G. Chen, T. Lang, J. He, "Birefringence compensated silicon nanowire arrayed waveguide grating for CWDM optical interconnects," Opt. Lett. **39**, 1834–1837 (2014).
32. T. Ye, T. Chu, "Low-loss and low-crosstalk Si etched diffraction gratings with multi-point iterative optimization," IEEE Int. Conf. Gr. IV Photonics (GFP), Shanghai, China, August, 2016.
33. R. J. Lycett, D. F. Gallagher, V. J. Brulis, "Perfect chirped echelle grating wavelength multiplexor: design and optimization," IEEE Photonics J. **5**, 2400123–2400123 (2013).
34. S. Xie, Y. Meng, J. Bland-Hawthorn, S. Veilleux, M. Dagenais, "Silicon nitride/silicon dioxide echelle grating spectrometer for operation near 1.55 μm," IEEE Photon. J. **10,** 1–7 (2018).
35. P. Pottier, M. J. Strain, M. Packirisamy, "Integrated Microspectrometer with Elliptical Bragg Mirror Enhanced Diffraction Grating on Silicon on Insulator," ACS Photonics **1**, 430-436 (2014).
36. J. Song, N. Zhu, J.J. He, S. He, "Etched diffraction grating demultiplexers with large free-spectral range and large grating facets," IEEE Photonics Technol. Lett. **18**, 2695–2697 (2006).
37. Y. Zhao, C. Qiu, A. Wu, H. Huang, J. Li, Z. Sheng, W. Li, X. Wang, F. Gan, "Broadband polarization splitter-rotator and the application in WDM receiver," IEEE Photonics J. **11**, 1–10 (2019).